\documentclass[11pt, a4paper]{article}
\pdfoutput=1
\usepackage{jcappub}
\usepackage{amsmath}
\usepackage{amssymb}
\usepackage{graphicx}
\usepackage[mathscr]{euscript}
\usepackage{mathrsfs}
\usepackage{calligra}
\usepackage[toc,page]{appendix}
\usepackage[utf8]{inputenc}

\DeclareMathAlphabet{\mathpzc}{OT1}{pzc}{m}{it} 
\DeclareMathAlphabet{\mathcalligra}{T1}{calligra}{m}{n}
\DeclareFontShape{T1}{calligra}{m}{n}{<->s*[2.2]callig15}{}

\newcommand{\x}{\ensuremath{\mathcalligra{x}}}
\newcommand{\y}{\ensuremath{\mathcalligra{y}}}
\newcommand{\z}{\ensuremath{\mathcalligra{z}}}
\newcommand{\rr}{\ensuremath{\mathcalligra{r}}}

\newcommand{\br}{\ensuremath{\pmb{\mathcalligra{r}}}}
\newcommand*{\rrhat}[1]{#1\kern-0.32em\hat{\phantom{#1}}}
\newcommand*{\xxhat}[1]{#1\kern-0.39em\hat{\phantom{#1}}}
\newcommand*{\yyhat}[1]{#1\kern-0.52em\hat{\phantom{#1}}}
\newcommand*{\zzhat}[1]{#1\kern-0.46em\hat{\phantom{#1}}}

\newcommand{\xhat}{\ensuremath{\pmb{\xxhat{\mathcalligra{x}}}}}
\newcommand{\yhat}{\ensuremath{\pmb{\yyhat{\mathcalligra{y}}}}}
\newcommand{\zhat}{\ensuremath{\pmb{\zzhat{\mathcalligra{z}}}}}
\newcommand{\xp}{\ensuremath{\mathcalligra{x}^{\: \prime}}}
\newcommand{\yp}{\ensuremath{\mathcalligra{y}^{\: \prime}}}
\newcommand{\zp}{\ensuremath{\mathcalligra{z}^{\: \prime}}}
\newcommand{\rp}{\ensuremath{\mathcalligra{r}^{\: \prime}}}

\newcommand{\brp}{\ensuremath{\pmb{\mathcalligra{r}}^{\: \prime}}}
\newcommand{\xphat}{\ensuremath{\pmb{\xxhat{\mathcalligra{x}}}^{\prime}}}
\newcommand{\yphat}{\ensuremath{\pmb{\yyhat{\mathcalligra{y}}}^{\prime}}}
\newcommand{\zphat}{\ensuremath{\pmb{\zzhat{\mathcalligra{z}}}^{\prime}}}

\title{The kinetic Sunyaev-Zel'dovich effect from the diffuse gas in the Local Group}

\author[a]{Douglas Rubin}
\author[b]{and Abraham Loeb}

\affiliation[a]{Department of Physics, Harvard University, \\
Cambridge, MA 02138, USA}
\affiliation[b]{Department of Astronomy, Harvard University, \\
Cambridge, MA 02138, USA}

\emailAdd{dsrubin@physics.harvard.edu}
\emailAdd{aloeb@cfa.harvard.edu}

\abstract{Since the Local Group (LG) of galaxies moves with a bulk velocity with respect to the cosmic microwave background radiation (CMB), free electrons in its gaseous halo should imprint large-scale non-primordial temperature shifts in the CMB via the kinetic Sunyaev-Zel'dovich (kSZ) effect.  By modeling the distribution of gas in the LG halo and using its inferred velocity with respect to the CMB, we calculate the resulting kSZ signal from the diffuse LG medium.  We find that it is dominated by a hot spot $\sim 10^\circ$ in size in the direction of M31, where the optical depth of free electrons is the greatest.  By performing a correlation analysis, we find no statistical evidence that the kSZ signal from model of the LG halo is embedded in the CMB temperature map measured by the Planck satellite.  We constrain the amount of mass in the LG halo by limiting the kSZ temperature shift around the hot spot to be smaller than the observed temperature shift in the Planck map.  We find the tightest constraints for models where the halo mass is highly concentrated, with the mass limited to roughly $2.5-5\times 10^{12}$M$_\odot$, but note that halos with such high concentrations are rare.}


\begin{document}
\maketitle
\flushbottom

\section{Introduction}

The kinetic Sunyaev-Zel'dovich effect occurs when cosmic microwave background photons scatter off free electrons with bulk velocities relative to the cosmic rest frame \cite{sunyaev:1980, hogan:1992}.  The CMB temperature shift associated with the kSZ effect depends upon the distribution of free electrons along the line of sight and their radial velocities relative to the CMB, and is given by:
\begin{equation}
\label{eq:sz_eqn}
\frac{\Delta T}{T} = -\frac{\sigma_T}{c}\int_{\ell os} v_{\ell os} n_e d \ell.
\end{equation}
In this equation, $\sigma_T$ is the Thomson cross section, $c$ is the speed of light, $v_{los}$ is the line-of-sight velocity, $n_e$ is the electron number density and $\ell$ is the position variable along the line-of-sight.  Unlike the thermal SZ effect, the kSZ effect only induces a temperature shift, and does not alter the spectral shape of the CMB signal.  It is thus particularly difficult to remove kSZ contaminates from measurements of the primordial temperature fluctuations of the CMB.

Since measuring the power spectrum of the CMB temperature fluctuations is now a precision science, it is therefore worth the effort to understand foreground contaminants which induce even relatively small temperature shifts.  One such contaminant, which can produce temperature shifts of order a few $\mu$K  \cite{birnboim:2009, peiris:2010, haijan:2007}, is the kSZ shift due to free electrons in the local universe.  Additionally, these temperature shifts may help explain well known anomalies in the CMB sky maps.  Large scale anomalies such as the observed hemispherical asymmetry \cite{planck1, eriksen:2004} in the power spectrum amplitude are contrary to the expected statistical isotropy of the CMB signal.  Without resorting to exotic physics (and barring observation or analysis issues), one potential explanation is the kSZ effect from local sources, which, because of their proximity, produce large scale temperature shifts on the sky.

The kSZ effect due to free electrons in the local universe has been studied by several groups.  The expected kSZ signal from the Milky Way (MW) halo, inflowing filaments and high velocity clouds within the halo has been calculated in Ref. \cite{birnboim:2009}.  A cross correlation analysis with the WMAP5 data, however, showed no significant correlation.  The kSZ signal from the MW halo has also been considered by Ref. \cite{peiris:2010}, but in the context of attempting to explain the large scale anomalies in the CMB data.  Using different statistical metrics, they found that the kSZ signal from the halo could explain the observed anomalies, but only if the column density of free electrons is at least an order of magnitude larger than indicated by observations.  The kSZ signal from the MW itself has also been calculated by Ref. \cite{haijan:2007} by using the electron distribution inferred from pulsar dispersion measurements.  Both the kSZ and thermal SZ signals from local superclusters were calculated by Ref. \cite{dolag:2005}.  They simulated the local distribution of gas by using hydrodynamical simulations with the initial conditions constrained to reproduce prominent structures in the local universe.  They found that it could be possible to extract and estimate of the SZ signal at the largest scales from Planck's measurement of the CMB.

In this paper, we consider the kSZ effect from another local source: the diffuse intragroup medium associated with the Local Group (LG) of galaxies.  Although it has been difficult to detect observationally \cite{maloney:1999}, a gaseous halo surrounding groups of galaxies is predicted theoretically.  In particular, the mass of the LG medium ($\sim 10^{12}$M$_\odot$) is expected to be a substantial fraction of the total mass of the LG \cite{cox:2008}.  Assuming that the fraction of baryons in the LG medium follows the cosmic mean (about 17\%), the baryonic mass in the LG medium should therefore be $\sim10^{11}$M$_\odot$.  For comparison, after subtracting the total baryonic mass in stars and gas in the Galaxy, the diffuse baryonic mass of the MW halo is about $5 \times 10^{10}$M$_\odot$ \cite{birnboim:2009, peiris:2010}.  Since the total mass in diffuse gas in the LG medium is several times larger than that of the MW halo, it is possible that the kSZ temperature shift from the LG medium could be several times larger, or $ \sim 10\mu$K.  Indeed, the thermal SZ effect due to the diffuse LG medium was considered by Ref. \cite{suto:1996} and it was found that the temperature shift from the quadrupole term could be a considerable contaminant to the CMB.

The outline of this paper is as follows.  In \S 2, we present our model for the distribution of gas in the LG halo.  Using this model, we derive the formalism to actually calculate the kSZ signal in \S 3.  In \S 4, we present the sky maps and power spectra from the kSZ signal due to the LG medium.  In the same section we describe our correlation analysis of the kSZ maps with the Planck satellite map to test if the kSZ signal is embedded in the measured CMB temperature fluctuations.  In \S 5 we compare the kSZ maps in order to the Planck map to limit the amount of mass in the LG halo.  We conclude in \S 6 and discuss how our results might change by using more realistic models of the gas distribution in the LG.

\section{Physical model of the diffuse local group medium}
\label{sec:phys_model}

\begin{figure}
\centering
\includegraphics[clip, width=3.5in]{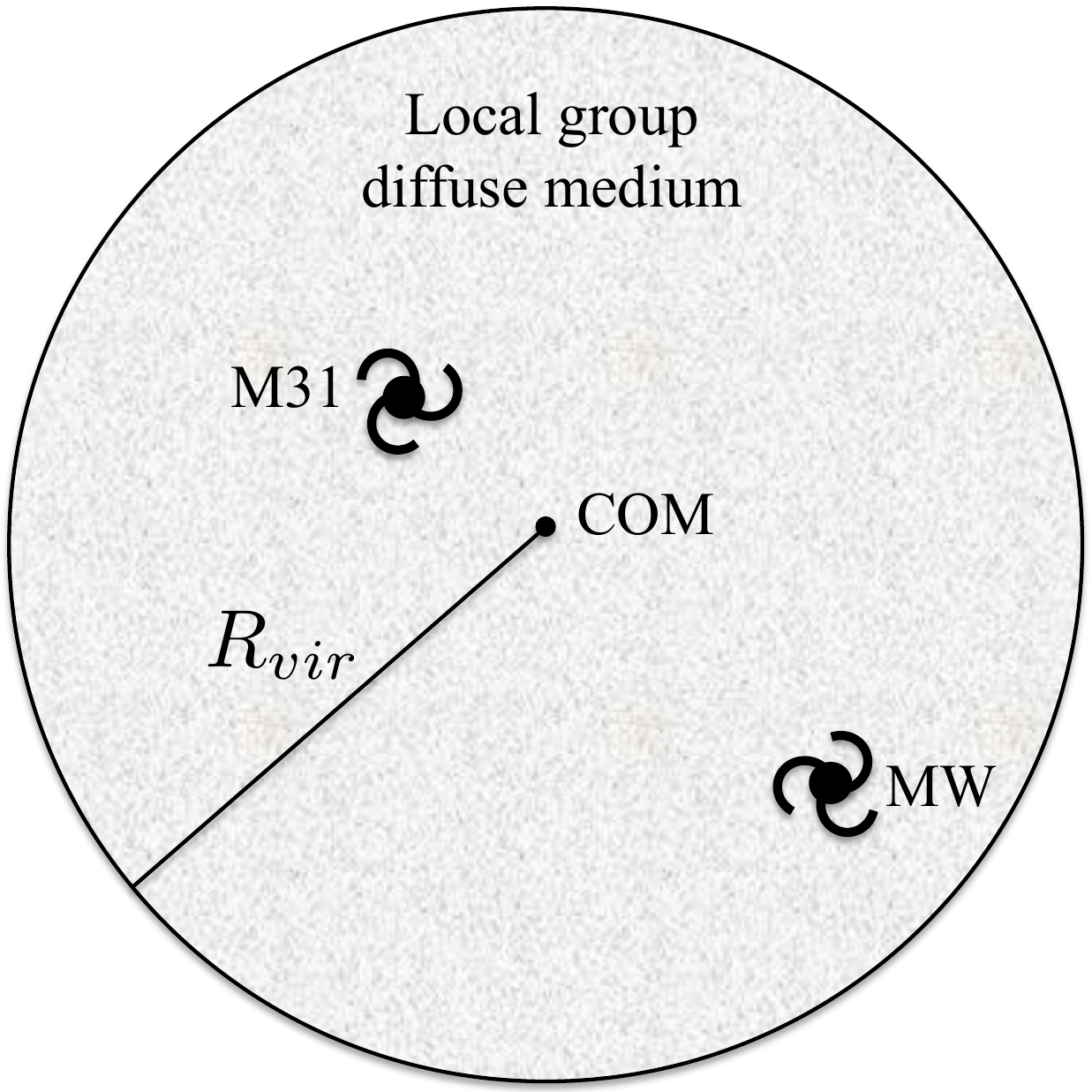}
\caption{ \label{fig:local_group_halo} An illustration of our model for the distribution of mass in the LG medium.} 
\end{figure}

In order to compute the kSZ temperature shift due to free electrons in the LG halo, we require a model of the baryonic content in the diffuse LG medium.  Several groups have modeled the distribution of mass in the LG medium in order to set the gravitational potential for various studies, such as simulating the collision between the MW and M31 (\cite{cox:2008}) and calculating the trajectories of hypervelocity stars (\cite{sherwin:2008}).  The LG medium was also modeled by Ref. \cite{suto:1996} to calculate the thermal SZ effect, and we follow their general approach by modeling the LG halo as a virialized sphere whose center coincides with the center of mass position of the galaxies in the LG.  The virial radius of the sphere, $R_{vir}$, can be calculated once the halo mass and virialization redshift are set.  An illustration of our model of the LG medium is shown in Fig.~\ref{fig:local_group_halo}.  Since the total mass in galaxies in the LG is dominated by the MW and M31, we only consider these two galaxies in determining the center of mass position of the LG galaxies.  The center of the virialized sphere will therefore be at a point on the line that passes through the centers of the MW and M31.

We adopt an NFW profile \cite{navarro:1996},
\begin{equation}
\label{eq:nfw}
\rho(R) = \frac{\rho_o}{\frac{R}{R_{vir}}\left (1+c\frac{R}{R_{vir}}\right )^2},
\end{equation}
as the density profile of the LG halo, where $R$ is the radial distance from the halo center.  The quantity, $c$, is the so-called concentration parameter, and $\rho_o$ is a normalization constant which can be re-expressed in terms of the virial mass by integrating the density profile out to the virial radius:
\begin{equation}
\rho_o= \frac{M_{vir}c^2}{4 \pi R_{vir}^3 \left [\ln(1+c)-\frac{c}{1+c}\right]}.
\end{equation}
Once a halo's mass and virialization redshift are specified, its virial radius can be found from
\begin{equation}
R_{vir} \cong 1.5 \mathrm{kpc} \left [\frac{\Omega_m}{\Omega_m(z_{vir})} \frac{\Delta_c(z_{vir})}{18 \pi^2} \right ]^{-1/3} \left (\frac{M_{vir}}{10^8 M_\odot} \right)^{-1/3} \left (\frac{1+z_{vir}}{10} \right)^{-1},
\end{equation}
where the pre-factor of 1.5 was calculated using $h=0.7$ (the Hubble constant in units of 100km/s/Mpc), and where $\Delta_c (z_{vir})$ is the mean density at collapse in units of the critical density of the universe at collapse.  This quantity is typically found from the spherical collapse model, and for a flat universe with a cosmological constant is well fit by
\begin{equation}
\Delta_c = 18 \pi^2 +82 d - 39 d^2,
\end{equation}
with $d \equiv \Omega_m(z_{vir})-1$ \cite{loeb:2012}.

Although there have been several studies to determine the total mass of the LG group (\cite{vm:2008, li:2008, vm:2012}), the exact amount of mass within the LG medium is still uncertain.  We therefore parameterize $M_{vir}$ as
\begin{equation}
M_{vir} = \eta (m_{mw}+m_{M31}),
\end{equation}
with $\eta$ as a free variable.  As in Refs. \cite{cox:2008} and \cite{sherwin:2008} we take $m_{M31} = 1.6 \times 10^{12}\mathrm{M}_{\odot}$ and $m_{MW} = 10^{12}\mathrm{M}_{\odot}$, consistent with a range of observations and simulations (see Ref. \cite{vm:2012} for a review on the literature of the masses of the MW and M31).  If the amount of mass in the LG medium is is equal to the combined mass of the MW and M31 (as assumed by \cite{cox:2008}), $\eta=2$, and assuming that the LG has virialized only recently ($z_{vir}=0$) we find that $R_{vir} = 402.6$kpc and $\rho_o =6.27 \times 10^4$M$_\odot$kpc$^{-3}$.  These values were calculated with $c=4$, a typical concentration parameter for recently formed halos \cite{zhao:2009}.

To model the density profile of the diffuse gas in the halo, we assume that the distribution of baryons follows the dark matter and that the fraction of baryons to dark matter in the LG medium is the same as the cosmic average, $f_b\equiv\Omega_b/(\Omega_b+\Omega_{DM})\cong0.15$.  In this case, the density profile of the baryons is found by multiplying Eqn.~\ref{eq:nfw} by the fraction, $f_b$:
\begin{equation}
\rho_b(R) = \frac{f_b \rho_o}{\frac{R}{R_{vir}}\left (1+c\frac{R}{R_{vir}}\right )^2}
\end{equation}
With our model for the baryonic density profile set, and with all the quantities in this model ($\rho_o$, $R_{vir}$, $M_{vir}$) found from the expressions provided above, we may now determine the number density profile of free electrons for use in calculating the kSZ temperature shift.  Assuming that the relative fraction of hydrogen to helium in the LG is primordial,
\begin{equation}
n_H \approx 12 n_{He},
\end{equation}
and that the mass from heavier elements is negligible, the baryon mass density is given by
\begin{equation}
\rho_b = m_p n_H+4 m_p n_{He}.
\end{equation}
In the above equation we have used the highly accurate approximation that the mass of a neutron is equal to the mass of a proton.  Since the gas temperature in the LG halo is $\sim 10^6 - 10^7$K, hydrogen and helium are fully ionized so that 
\begin{equation}
n_e = n_H +2 n_{He}.
\end{equation}
Using the previous three equations, we find that the free electron density profile in the diffuse LG medium is given by
\begin{equation}
n_e(R) \cong 0.88 \left[ \frac{\rho_b(R)}{m_p} \right].
\end{equation}

\section{Calculating the kSZ signal}
\label{sec:geom}

In order to calculate the kSZ temperature shift due to the diffuse LG medium, we must integrate the free electron density along a particular line of sight (parameterized by $\bold{\hat r}(\theta, \phi)$) from our observation point near the Sun.  However, since the free electron density is expressed as a function of radial distance away from the center of mass origin, $R$, we must convert a given radial distance from the Sun, $r$, (for a particular $\bold{\hat r}$ direction) to a value of $R$.  In this section we go over the geometry needed to accomplish this, but leave lengthy derivations for the Appendix.

\begin{figure}
\centering
\includegraphics[clip, width=5.5in]{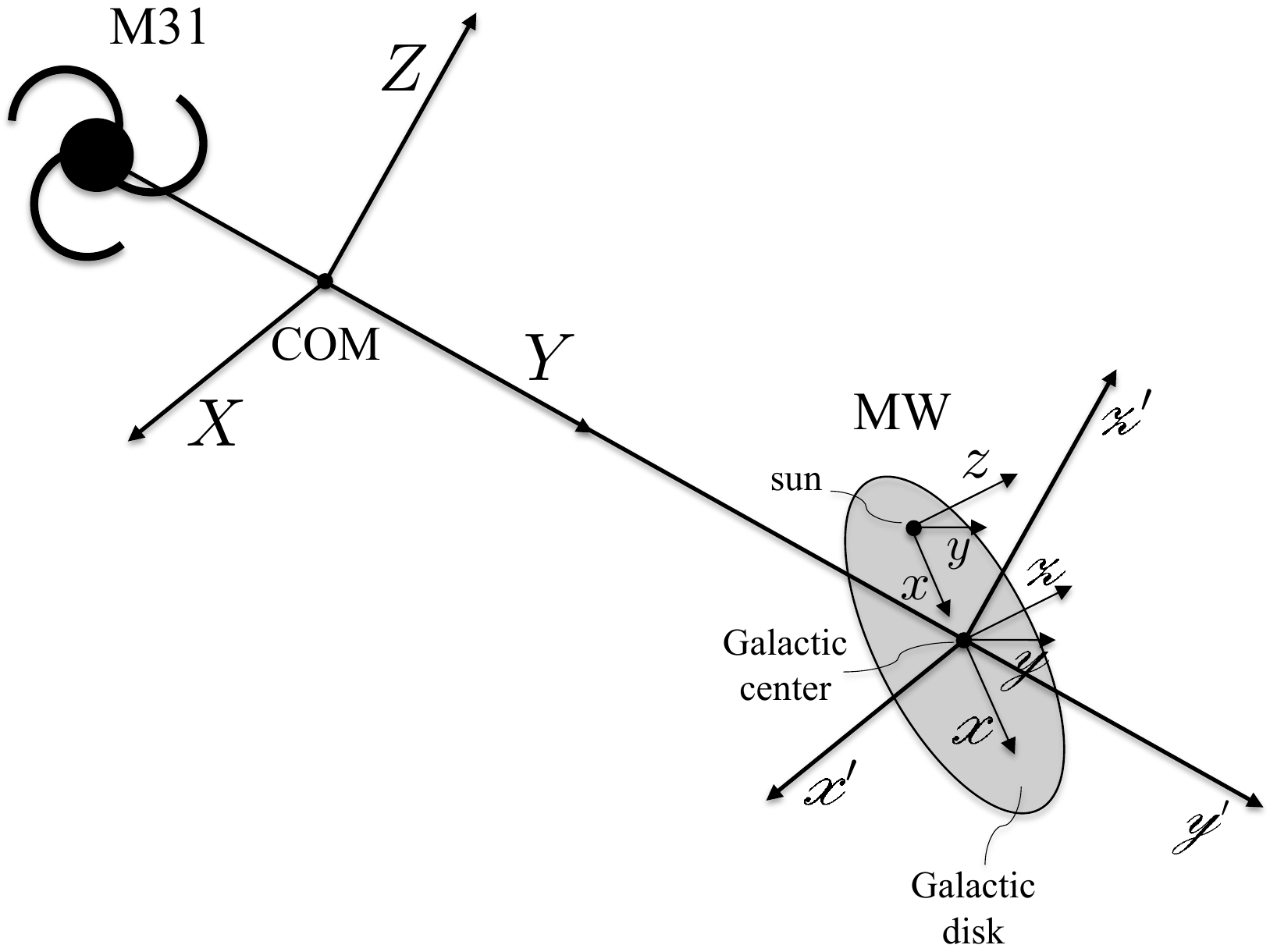}
\caption{ \label{fig:geometry} The coordinate systems used to convert from the heliocentric to center of mass frames.} 
\end{figure}

To derive the geometric conversion, we define several cartesian coordinate systems shown in Fig.~\ref{fig:geometry}.  We place a coordinate system, denoted by ($X$, $Y$, $Z$), at the center of mass position and orient the axes such that its $Y$ axis points directly toward the Galactic center.  Another coordinate system, denoted by ($\xp$, $\yp$, $\zp$) is placed at the Galactic center and we orient its axes to be in the same direction as the center of mass system.  We set the $\xp$ axis to be in the Galactic plane.  We define a heliocentric coordinate system, denoted by ($x$, $y$, $z$), in the usual way by placing the origin at the sun and orienting the $x$ and $y$ axes in the Galactic plane with the $x$ axis pointing toward the Galactic center.  This coordinate system is commonly called the Galactic coordinate system.  We also place a coordinate system at the Galactic center whose axes, denoted by ($\x$, $\y$, $\z$), are oriented in the same way as the heliocentric system (a Galactocentric Galactic coordinate system).  To aide the reader for the equations in the rest of this paper, the notation associated with each coordinate system is shown in Table~\ref{tab:coord_notation}.  From Fig.~\ref{fig:geometry}, it can be seen that any position vector in the Galactic coordinate system can be expressed in the center of mass system by a translation to the Galactocentric frame, a rotation to the ``primed" frame and another translation to the center of mass frame.  In Appendix \ref{sec:convert} we perform these operations to express R as a function of $r$, $\theta$ and $\phi$, and in Appendix \ref{sec:rotation}, we derive the rotation matrix between the primed and Galactocentric frames.

  \begin{table}[h]
\caption{Coordinate system notation in the Local Group} 
\centering{ 
\begin{tabular}{ll} 
\hline\hline 
coordinate system & notation
\\ [0.5ex]
\hline
  \\[-2.ex]
heliocentric frame & $(x, y, z, r, \theta, \phi)^{\mathrm{a}}$ \\
Galactocentric frame & $(\x, \y, \z, \rr)$ \\
primed frame & $(\xp, \yp, \zp, \rp)$\\
center of mass frame & $(X, Y, Z, R)$ \\ [,5 ex]
\hline
\multicolumn{2}{l} {$\scriptstyle{\mathrm{^aFor~the~derivations~presented~in~this~paper~we~only}}$} \\[-1.2ex]
\multicolumn{2}{l} {$\scriptstyle{\mathrm{require~angular~variables~for~the~heliocentric~frame.}}$}
\end{tabular}}
\label{tab:coord_notation}
\end{table}

To perform these operations, we also require several observed quantities.  We take the distance from the sun to the Galactic center to be $d_{\odot} = 8.3$kpc, consistent with Refs. \cite{gillessen:2009, mcmillan:2011}.  To find the direction to the center of mass, we use the position vector of M31 (in the Galactocentric coordinate system) given by \cite{vm:2012},
\begin{equation}
\label{eq:r_m31}
\br_{M31} = (-378.9, 612.7, -283.1)\mathrm{kpc},
\end{equation}
so that the distance from the Galactic center to M31 is 
\begin{equation}
d_{M31}= |\br_{M31}| = 774.0\mathrm{kpc}.
\end{equation}
Since we ignore the contributions from dwarf satellites in calculating the center of mass position of the LG galaxies, the distance from the Galactic center to the center of mass is
\begin{equation}
d_{com} = d_{M31} \left (\frac{m_{M31}}{m_{M31}+m_{mw}}\right ) = 476.3\mathrm{kpc}.
\end{equation}
The center of mass position vector expressed in the Galactocentric frame is therefore the distance, $d_{com}$, times a unit vector in the direction of $M31$:
\begin{align}
\label{eq:com_position}
\br_{com} & = d_{com} \left (\frac{\br_{M31}}{d_{M31}} \right) \nonumber \\
& = (-233.2, 377.0, -174.2)\mathrm{kpc}.
\end{align}

Having defined the relevant geometry to calculate $R(r, \theta, \phi)$, we may now calculate the kSZ temperature shift from the LG medium.  To do this, we re-write Eqn.~\ref{eq:sz_eqn} as
\begin{equation}
\frac{\Delta T}{T}\left(\theta, \phi \right) = -\frac{1}{c} \left (\bold{v}_{LG-CMB} \cdot \bold{\hat r} \right) \tau \left(\theta, \phi \right),
\end{equation}
where $\bold{\hat r}=\bold{\hat r}(\theta, \phi) = (\sin \theta \cos \phi, \sin \theta \sin \phi, \cos \theta)$, so that
\begin{equation}
\bold{v}_{LG-CMB} \cdot \bold{\hat r} = v_{LG-CMB}^{x} \sin \theta \cos \phi +v_{LG-CMB}^{y} \sin \theta \sin \phi +v_{LG-CMB}^{z} \cos \theta,
\end{equation}
and where $\tau$ is the optical depth along the line of sight.  From Ref. \cite{loeb:2008}, the velocity of the local group with respect to the CMB in the heliocentric frame is
\begin{equation}
\bold v_{LG-CMB} = (-1.8, -537.2, 293.2)\mathrm{km\, s^{-1}}.
\end{equation}

Since we are not situated at the center of mass origin, the optical depth depends on the particular line of sight.  By using Eqn.~\ref{eq:big_r} in Appendix~\ref{sec:convert} to re-write $R$ in terms of the heliocentric variables $r$, $\theta$ and $\phi$, the optical depth is given by integrating the density along the heliocentric radial direction:
\begin{equation}
\label{eq:tau_eqn}
\tau \left(\theta, \phi \right) = \sigma_T \int_0^{r_{vir}(\theta, \phi)}n_e \left ( \sqrt{r^2+d_\odot^2-2 d_\odot r \sin \theta \cos \phi +d_{com}^2 +2d_{com}\yp(r, \theta, \phi)} \right)dr.
\end{equation}
The $\yp$ value at a particular set of ($r$, $\theta$, $\phi$) is given in Eqn.~\ref{eq:yp}.  The integration runs till the edge of the virial sphere, at a distance $r_{vir}(\theta, \phi)$ away from the Sun.  This is found for a particular line of sight by evaluating Eqn.~\ref{eq:r_r} at $R=R_{vir}$,
\begin{align}
r_{vir}(\theta, \phi) \equiv r(\theta, \phi, R=R_{vir})=-B(\theta, \phi)+\sqrt{B^2(\theta, \phi)+C(R_{vir})},
\end{align}
where the functions $B$ and $C$ are defined in Eqns.~\ref{eq:b} and \ref{eq:c}.

\section{The kSZ signal due to the diffuse local group medium and its correlation with the CMB}

\begin{figure}
\centerline{\mbox{  \includegraphics[clip, width=3.2in]{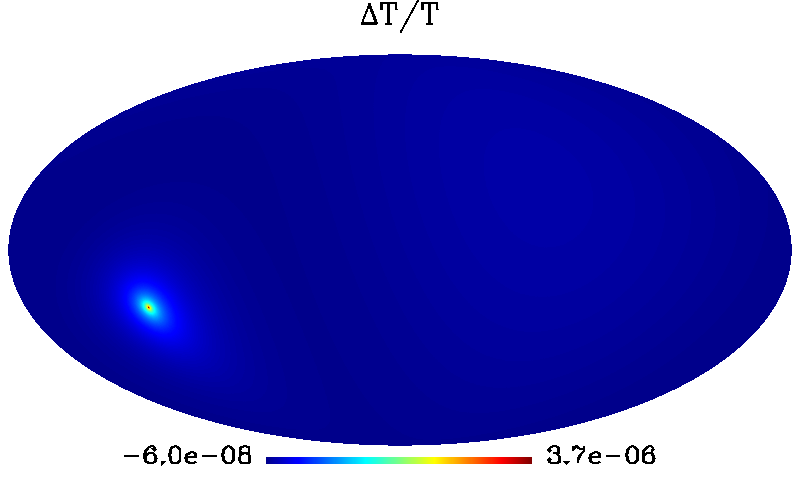}}
\mbox{  \includegraphics[clip, width=3.22in]{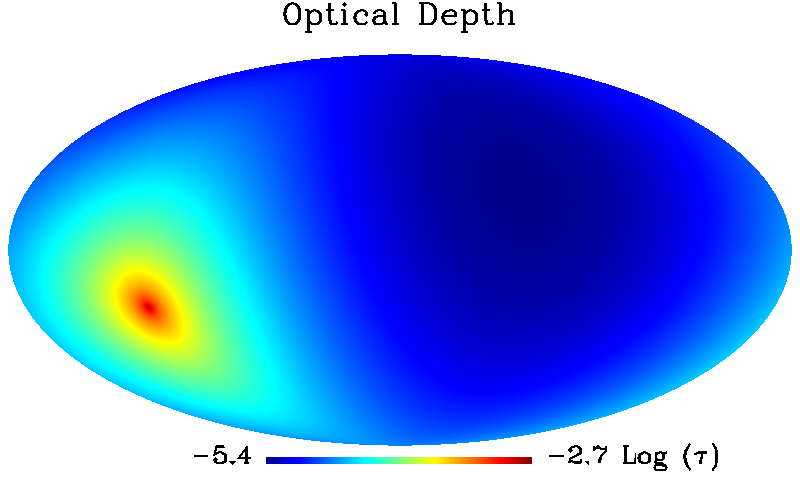}}}
\caption{ \label{fig:sky_maps}  Sky maps of the $\Delta T/T$ kSZ signal (monopole and dipole subtracted) and the optical depth from the diffuse LG medium for our fiducial model with $c=4$ and $\eta=2$.} 
\end{figure}

Using the methodology described in the previous section, we calculate the expected kSZ signal from the diffuse medium associated with the local group.  We choose a fiducial model of $\eta=2$ and $c=4$ and show a Mollweide projection in Galactic coordinates of the fractional temperature shift in the first panel of Fig.~\ref{fig:sky_maps}.  We have subtracted the monopole and dipole components of the map.  We use the HEALPix pixelization scheme \cite{gorski:2005} with Nside = $2^9$.  In the second panel we show a map of the optical depth calculated with Eqn.~\ref{eq:tau_eqn}.  The figure shows that the temperature shift due to the gaseous halo of the local group is dominated by a hot spot several degrees in size.  The spot is in the direction of M31 and is due to the fact that in our model the center of mass position is in the direction of M31.  From Fig.~\ref{fig:local_group_halo} it is evident that not only does this direction correspond to the greatest path length through the medium, but it also passes directly through the center where the density is the highest.  This results in the greatest optical depth along that line of sight.  The scale shows that the temperature shift can be order several $\mu$K, which exceeds the detectability threshold of the CMB map measured by Planck.

\begin{figure}
\centering
\includegraphics[clip, width=5.5in]{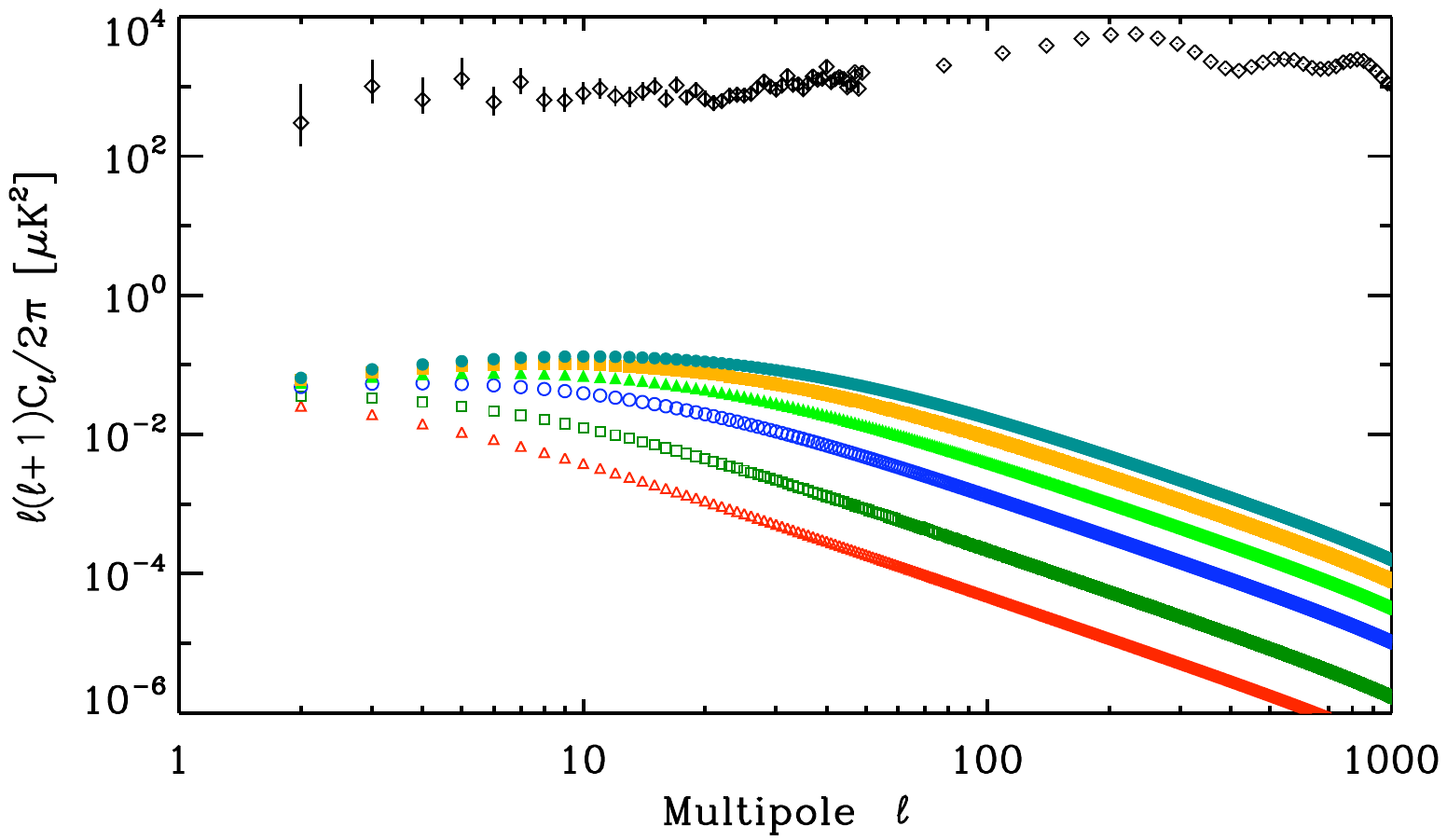}
\caption{ \label{fig:power_spectrum}  The angular power spectrum of our kSZ sky maps calculated with $c =$ 2, 4, 8, 12, 16 and 20 (red open triangles, green open squares, blue open circles, light-green filled triangles, orange filled squares and teal filled circles respectively).  The black open diamonds correspond to the angular power spectrum of the CMB temperature fluctuations as measured by Planck \cite{planck2}.} 
\end{figure}

We show the angular power spectrum of the kSZ signal for several values of $c$ in Fig.~\ref{fig:power_spectrum}.  The figure shows that most of the power comes from large scales (low $\ell$) consistent with the fact that the map is dominated by a large hot spot.  At the smallest scales (highest values of $\ell$), the amount of power has a clear dependence on $c$, with the highest values resulting in the greatest power.  We also show the measured power spectrum of CMB temperature fluctuations from Planck \cite{planck2}.  At the largest scales, the power from the temperature shift induced by the kSZ effect from the LG halo can be as much 0.01\% that of the primordial CMB temperature fluctuations.  This, of course, depends on value of $\eta$ used, which in this figure is $\eta=2$.  Since the overall amplitude of the kSZ temperature shift is proportional to $\eta$ ($\Delta T \propto n_e \propto \rho_o \propto M_{vir} \propto \eta$, as quickly verified from the equations presented above), the lines in this figure scale as $\eta^2$, and thus have a strong dependence on $\eta$.  However, as described in \S~\ref{sec:phys_model} a reasonable estimate of the amount of gas in the LG diffuse medium limits $\eta$ to be close to unity.

\begin{figure}
\centering
\includegraphics[clip, width=4.5in]{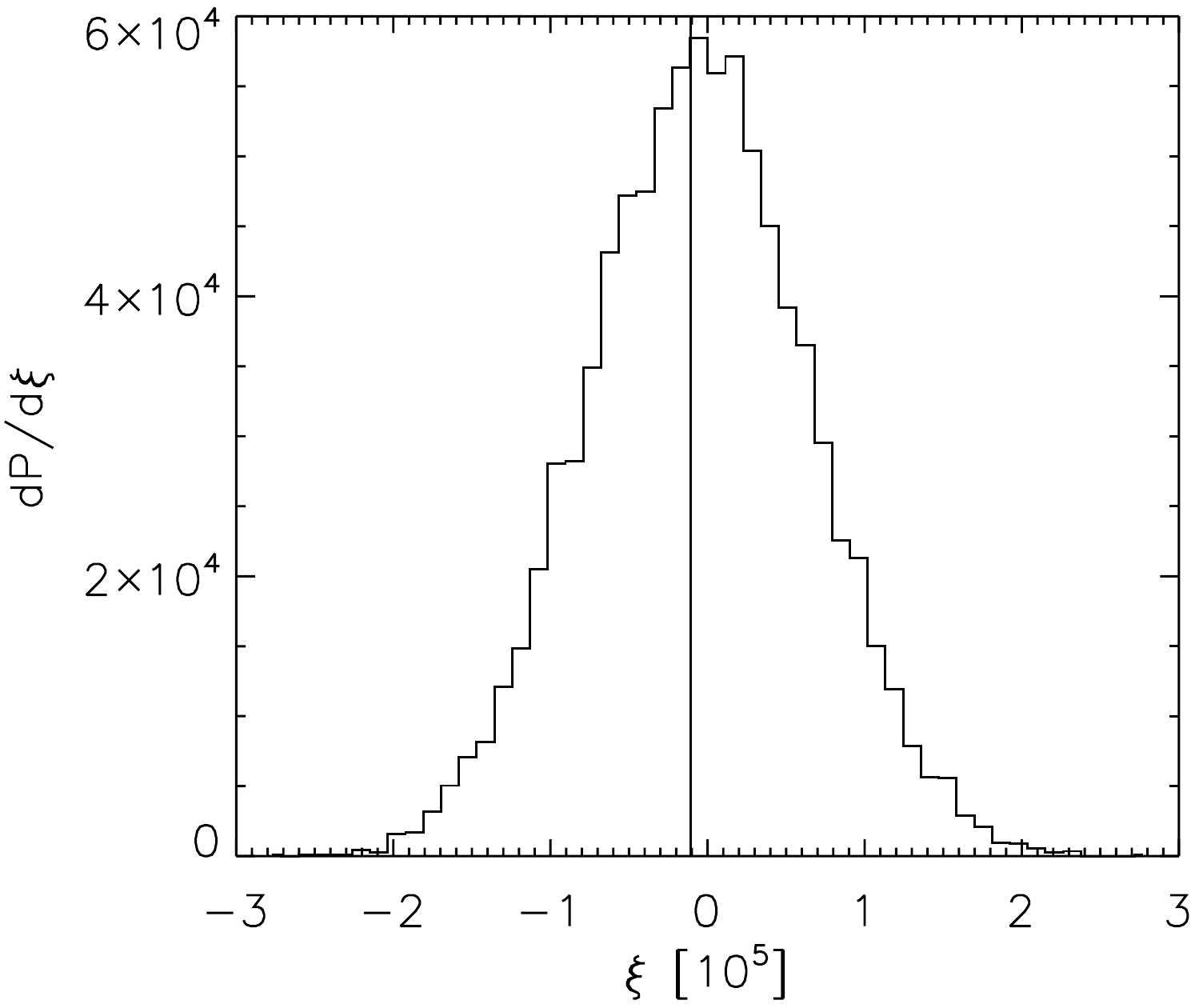}
\caption{ \label{fig:prob_dist} The distribution of $\xi$ generated from random realizations of the CMB power spectrum.  The vertical line gives the correlation from the temperature map measured by Planck.} 
\end{figure}

In order to test whether the kSZ signal from the diffuse LG medium is embedded in the CMB map measured by Planck, we perform the following correlation analysis.  We first rebin the (background/foreground subtracted) CMB temperature shift map from Planck from its original size of Nside = $2^{11}$ to Nside = $2^7$ in order to keep the analysis computationally manageable.  The mask is also rebinned to Nside = $2^7$, and all pixels with a value less than 1 are set to 0.  We then calculate the kSZ map from our fiducial model with the same Nside and subtract the monopole and dipole components.  We compute the correlation at zero separation between the two maps (excluding bad pixels as indicated by the rebinned mask), $\xi$ , with $\xi \equiv \langle \delta_{CMB} \delta_{kSZ} \rangle$.  Here, $\delta_{CMB} \equiv (T_{CMB}-T_o)/T_o = \Delta_{CMB}/T_o$, where $\Delta_{CMB}$ is the rebinned Planck map, and where $T_o =2.7255K$ and $\delta_{kSZ} \equiv (\Delta T/T-\overline{\Delta T/T})/\overline{\Delta T/T}$.  Given this definition of $\xi$, if no correlation exists between the two maps, then $\xi=0$.  To test the significance of this correlation value, we adopt a Monte Carlo approach where we make mock CMB maps from a given power spectrum and calculate the same correlation (again excluding the same bad pixels), but using the mock CMB maps instead of the Planck data.  We use the power spectrum calculated by CAMB using the best fit cosmological parameters from Planck \cite{planck3}.  To generate random realizations from the power spectrum, we use the HEALPix IDL procedure isynfast.pro, with a FWHM beamsize of 5 arcminutes (the beamsize of the Planck data \cite{planck1}).  We calculate $\xi$ for 10,000 realizations and show the results as a probability distribution in Fig.~\ref{fig:prob_dist}.  In this figure, the vertical line indicates the value of $\xi$ calculated from the original Planck map.
Not only is the value of the correlation calculated with the original Planck map slightly negative (anti-correlated), but it is clear from the figure that the standard deviation of $\xi$ is many times bigger than than this value.  It is therefore impossible to tell whether this slightly negative correlation is due to the actual kSZ effect or a chance realization of the primordial CMB signal, and we find no statistical evidence to suggest that the kSZ signal from the LG medium is embedded in the Planck map.  In order to test for possible resolution dependence, we have performed the same analysis for several values of Nside and have come to the same conclusion.

\section{Limiting the baryonic mass in the Local Group medium}

\begin{figure}
\centering
\includegraphics[clip, width=4in]{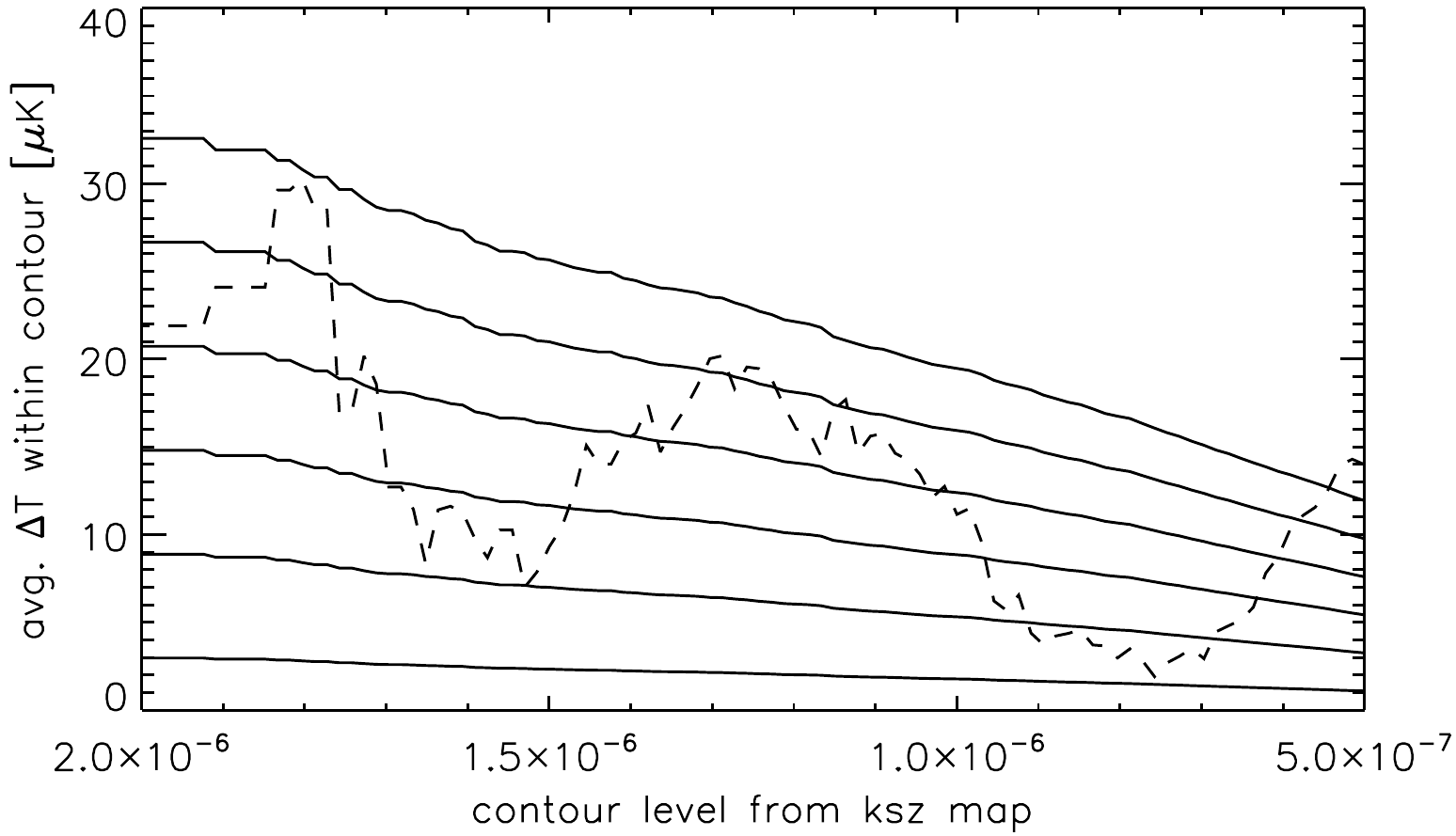}
\caption{ \label{fig:limit_eta}  The average temperature shift within a contour level plotted as a function of the contour level.  The levels, constructed from our fiducial kSZ map, are centered on the hot spot and extend out to about the size of the hot spot.  The dashed line denotes the average temperature shift from the Planck map, while the solid lines denote the average temperature shift from our kSZ maps for different values of $\eta$.  The solid lines extend from $\eta =1$ (lowest line) to $\eta=11$ (highest line) in steps of 2.} 
\end{figure}

Since the kSZ temperature shift due to the diffuse LG medium is concentrated in a hot spot in the direction of M31, we can place an upper limit on the amount of allowed mass in the LG medium.  This is done by ruling out halo masses which result in temperatures around the hot spot which are greater than what is actually observed in the CMB data.  To do this, we first make contour levels from our fiducial map extending from the brightest pixel in the hot spot to about the size of the hot spot.  We then calculate the average temperature shift within each of these contours for both the kSZ map and the Planck map.  We exclude bad pixels as indicated by the (rebinned) mask.  This is shown in Fig.~\ref{fig:limit_eta}, where have plotted the average temperature shift within a contour versus the contour level for both the kSZ signal for several values of $\eta$ (solid lines) and the Planck map (dashed line).  The horizontal axis is plotted from the highest contour to lowest contour which corresponds to the center of the hot spot at the origin.  The amplitude of the kSZ lines scale as $\eta$, since, as mentioned in the previous section, the kSZ shift is proportional to $\eta$.    The highest line, with $\eta \approx 11$, is clearly not allowed since it results in a temperature shift greater than observed.  We note that the maximum allowed value of $\eta$ depends slightly on the size of the maps used.  We use Nside = $2^7$ since this results in the tightest constraint on $\eta$.  Since $\eta$ represents the mass of the LG medium plus the masses of M31 and the MW in units of $m_{M31}+m_{MW}$, the total amount of allowed mass in the medium is $(\eta_{max}-1)(m_{M31}+m_{MW})\approx  10(m_{M31}+m_{MW})$.

\begin{figure}
\centering
\includegraphics[clip, width=4in]{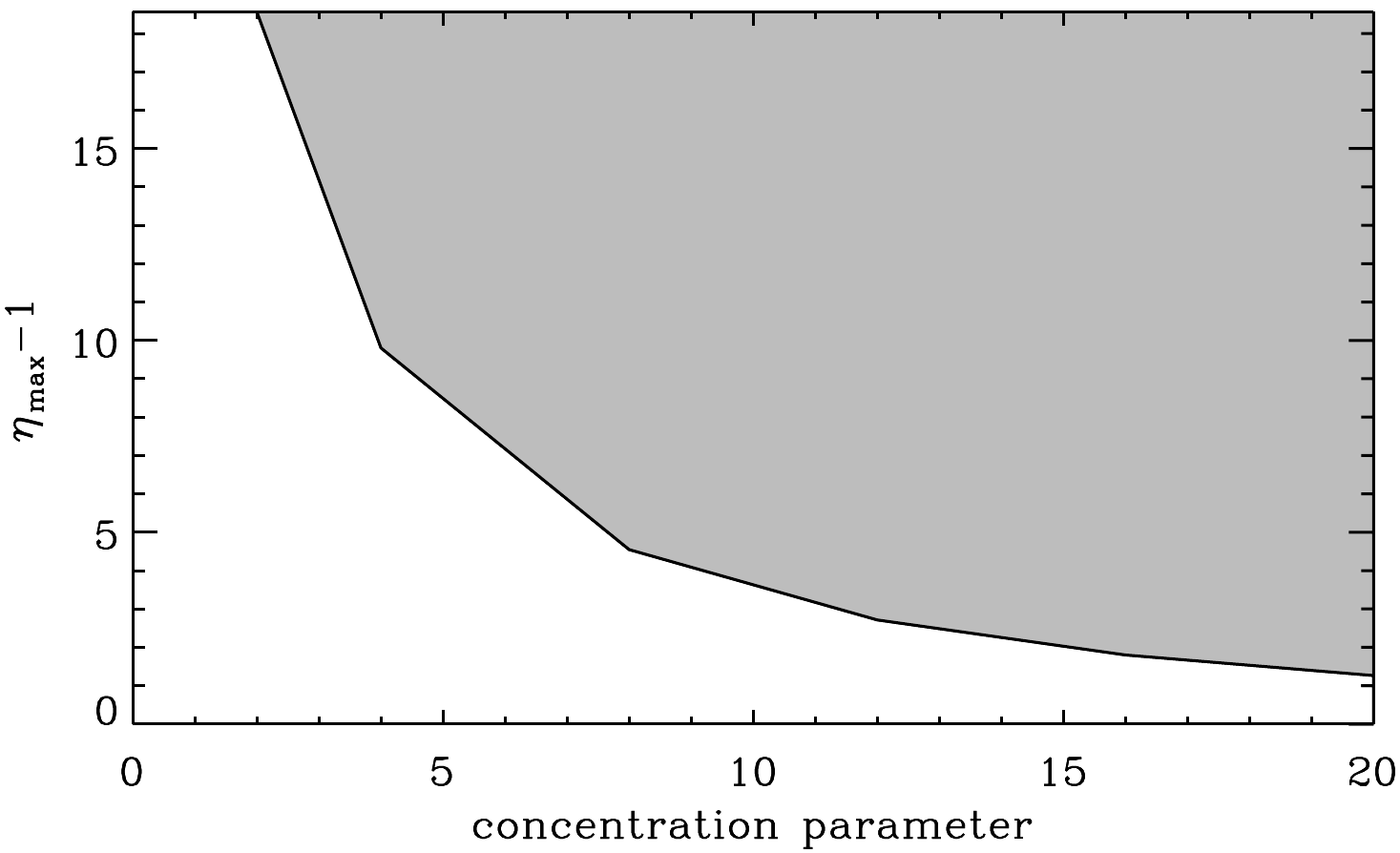}
\caption{ \label{fig:eta_max} The maximum amount of allowed mass in the LG halo (in units of $m_{M31}+m_{MW}$) as a function of the concentration parameter assumed for the LG halo profile.  The shaded region of the plot denotes values of $\eta_{max}-1$ not allowed by our analysis.} 
\end{figure}

Since the concentration parameter is a free parameter in our model, we repeat the same calculation for different values of $c$ to obtain the maximum allowed mass in the LG medium as a function of $c$.  We show in the results in Fig.~\ref{fig:eta_max}.  The shaded part of the figure denotes the region not allowed by our analysis.  By comparison, the so-called ``timing argument" estimates the amount of mass in the LG to be about $5\times 10^{12}$M$_\odot$ \cite{li:2008} (see also Ref. \cite{partridge:2013} who include the dynamical effects of dark energy in the timing argument).  The smallest values of $c$ result in upper mass limits much greater than the timing argument estimate, while the largest ($c \gtrsim 15$) result in mass limits roughly consistent with it.  

It has been found from numerical simulations that there exists a reasonably tight correlation between the concentration of a halo and its mass (where the halo mass is identified at a certain redshift) \citep[e.g.][]{bullock:2001, eke:2001, maccio:2008, zhao:2009, prada:2012}.  For a given redshift and halo mass, there is, of course, a spread in the concentration parameters measured.  For example, by analyzing the halos in the GIF2 simulation, \cite{giocoli:2012} find that, at $z=0$, for a halo mass roughly equal to that of the LG (a few $\times 10^{12} $M$_\odot$) the third quartile of the distribution corresponds to $c \approx 12$ (see their figure 9).  Using the Millennium Simulation, for the same mass and redshift, \cite{neto:2007} find a slightly lower value of $c\approx 10$ for the third quartile.  Disregarding complexities such as whether to consider all halos or only relaxed halos, in general it is found that, at a given mass and redshift, the distribution in $c$ is well fit by a lognormal function with variance $\sigma_{\mathrm{log} c} \cong 0.12 \pm 0.2$ \citep[][]{bullock:2001, wechsler:2002, dolag:2004, neto:2007}.  Depending on which fitting formula one uses to calculate the mean value of $\log(c)$ for a halo with a mass equal to that of the LG at $z=0$, the 3$\sigma_{\mathrm{log}c}$ value away from the mean corresponds to roughly $c\approx 15-20$.  It is therefore improbable that the concentration parameter of the LG is greater than 20, and we therefore truncate the x-axis of Fig.~\ref{fig:eta_max} at this value.  We do note, however, that even though we obtain the tightest constraints on $\eta_{max}$ for the highest values of $c$ in this figure, even these values of $c$ can be quite rare.

\section{Discussion and Conclusions}

We have modeled the distribution of baryonic mass in the LG medium to calculate the column density of free electrons along a particular line of sight.  Our model assumes that the mass in the halo is distributed as an NFW profile, that the baryons trace the dark matter and that the ratio of baryons to dark matter follows the cosmic average.  We have calculated sky maps of the possible kSZ signal due to the gaseous halo surrounding the LG, and have found that the it is concentrated in a hot spot several degrees in size in the direction of M31.  The maximal temperature shift is $\sim 10 \mu$K, several times bigger than the kSZ effect due to the MW and MW halo.  The kSZ signal, however, is still far sub-dominant to the primordial CMB temperature fluctuations as seen from its power spectrum.  Using random realizations of the CMB fluctuations, we found that the correlation of our kSZ signal with the Planck map is statistically insignificant.  By ruling out halo masses resulting in temperature shifts around the kSZ hot spot greater than observed, we were able to place an upper limit on the amount of mass in the LG medium.  We found that for the largest concentration parameters used to model the density profile of the LG medium, the amount of allowed mass is the most tightly constrained.  For the largest values of $c$, the mass in the halo is constrained to be about $\eta_{max}-1 \approx1-2$ (see Fig.~\ref{fig:eta_max}), corresponding to roughly $2.5-5\times 10^{12}$M$_\odot$ (where we have multiplied $(\eta_{max}-1)$ by $(m_{M31}+m_{MW})$ to convert to halo mass).  However, we again note that it is very rare for halos at $z=0$ with a mass comparable to that of the LG to have such large concentration parameters.

Our model of the diffuse LG medium assumes a spherically symmetric density profile centered on the center of mass position of M31 and the MW.  In reality, the distribution of gas will not have perfect spherical symmetry, will have small-scale spatial inhomogeneities and will have a center of mass position that might not coincide with the center of mass position of the LG galaxies.  It is possible that a more realistic model of the distribution of gas in the LG halo may change the significance of the correlation between the kSZ signal and the Planck map.  The upper limit on the amount of mass in the LG halo inferred from our kSZ sky maps could change as well.  However, as discussed below, it is difficult to make a more realistic model of the diffuse gas in the LG halo since the actual distribution of gas is relatively unconstrained.

Small-scale spatial inhomogeneities would lead to more textured kSZ sky maps and could significantly increase the power spectrum at larger values of $\ell$.  Indeed small-scale inhomogeneities have been included in studies of the kSZ signal due to the MW halo \cite{birnboim:2009} and local superclusters \cite{dolag:2005}.  The former study included a prescription for including high velocity clouds and infalling filaments of gas in their model of the gaseous halo of the MW.  These structures are seen clearly in their kSZ sky maps as small-scale inhomogeneities in the kSZ signal.  The inclusion of these features drastically enhances the angular power spectrum of the kSZ signal at $\ell \gtrsim 10$, compared to the case where only a smooth, gas density profile is included.  Still, at the highest multipoles, the kSZ power spectrum is subdominant to the WMAP5 primordial power spectrum by several orders of magnitude.  The latter study calculated the kSZ signal from hydrodynamic simulations, constrained to reproduce prominent observed structure in the local universe.  Small-scale inhomogeneities in the simulation lead to a roughly monotonic increase with $\ell$ in their measured kSZ power spectrum.  This is in contrast to our kSZ power spectrum, which shows a monotonic decrease with $\ell$ (Fig.~\ref{fig:power_spectrum}).  Again, even at the highest values of $\ell$, their power spectra are subdominant to the primordial CMB spectrum.  We therefore do not expect the inclusion of small-scale gaseous structure in our model of the diffuse gas in the LG to significantly enhance the kSZ power spectrum (compared to the primordial signal).  It is also not probable that the inclusion of small scale structure will lead to a significant correlation with the CMB signal measured by Planck, as this is highly dependent on precisely how this structure is distributed on the sky.  Indeed, the distribution of small-scale structure in the LG is relatively unconstrained, and it would therefore be difficult to even devise a prescription for the inclusion of gaseous inhomogeneities. 

Numerical simulations have shown that dark matter halos are in general not spherically symmetric, but are instead better approximated as ellipsoidal.  An ellipsoidal geometry, however, is more difficult to model as it has more free parameters (the ratio of the length of the axes as well as the orientation) which may not be well constrained.  One study actually has considered the kSZ signal due to an ellipsoidal gaseous halo associated with the MW \cite{peiris:2010}.  The authors of this study used axis ratios proposed by \cite{law:2009} in order to explain the observed properties of the Sagittarius dwarf spheroidal and considered several orientations.  Their main results, however, were not qualitatively different than as compared to a spherical geometry.  Any ellipsoidal model of the LG would be relatively unconstrained since the axis ratios and orientation for a triaxial model of LG halo are not well known.  

More complex models of the gaseous halo of the LG which could include small-scale inhomogeneities and triaxiality would have to cover a large parameter space since the actual distribution of gas in the LG halo is not well known.  We have therefore taken the simplest approach by modeling the gaseous halo with spherical symmetry.  This should reproduce the main features of the kSZ signal, specifically the presence of a large hot spot in the direction of M31.


\begin{appendices}

\section{Rotating between the Galactocentric and primed frames}
\label{sec:rotation}

The Galactocentric ($\x$, $\y$, $\z$) and primed ($\xp$, $\yp$, $\zp$) frames share the same origin, but their axes are rotated relative to each other (see Fig.~\ref{fig:geometry}).  In this section, we derive the rotation matrices for transforming a position vector from one frame to the other.  This can be done by solving a  coupled set of non-linear equations to solve for the Euler angles for the proper rotation matrix.  We, however, prefer to solve for the rotation matrix by hand, and present that derivation. 

By definition, the $\yphat$ unit vector points in the direction opposite M31, so that $\yphat$ can be written down immediately by normalizing the opposite of Eqn.~\ref{eq:r_m31}, $\yphat=-\br_{M31}/d_{M31}$, resulting in
\begin{equation}
\label{eq:y_hat}
\yphat = 0.4895\xhat-0.7916\yhat+0.3658\zhat.
\end{equation}
We are now free to orient the $\xp$ axis in which ever direction we please, so long as $\yphat$ and $\xphat$ are orthonormal.  For convenience, we choose to keep the $\xp$ axis in the Galactic plane.  In this case, $\xphat$ is written out as a linear combination of the $\xhat$ and $\yhat$ unit vectors: $\xphat = A\xhat+B\yhat$.  We solve for the coefficients $A$ and $B$ by noting that due to orthonormality $0.4895A-0.7916B = 0$ (since $\yphat \cdot \xphat=0$) and that $A^2+B^2=1$.  Solving these equations results in $B=\pm 0.5260$ and $A=\pm 0.8505$.  The 2 solutions are due to the fact that there are 2 unit vectors in the galactic plane, exactly opposite each other, which are orthonormal to $\yphat$.  The solution we choose will determine the direction of the $\zphat$ unit vector, since its direction is restricted by requiring a right handed coordinate system.  We choose the minus solution as this results in a $\zphat$ unit vector which points away from the galactic plane in the same direction as the $\zhat$ unit vector:
\begin{equation}
\label{eq:x_hat}
\xphat =  -0.8505\xhat -0.5260 \yhat.
\end{equation}
We solve for $\zphat$ by noting that in a right handed coordinate system, $\zphat=\xphat \times \yphat$, so that
\begin{align}
\label{eq:z_hat}
\zphat & =  \begin{vmatrix} \xhat & \yhat & \zhat \\
-0.8505 &  -0.5260 & 0 \\
0.4895 & -0.7916 & 0.3658
\end{vmatrix} \nonumber \\
& = -0.1924\xhat+0.3111\yhat+0.9307\zhat.
\end{align}
Indeed, since the $\zhat$ compnent is positive, $\zphat$ sticks above the galactic plane, as a consequence of us choosing the minus root solution of the $\xphat$ unit vector.

A position vector can be expressed in the primed basis as $\xp \xphat + \yp \yphat +\zp \zphat$ or in the Galactocentric basis as as $\x \xhat + \y \yhat +\z \zhat$.  By equating the two expressions, replacing the primed unit vectors in the former expression with Eqns.~\ref{eq:y_hat} - \ref{eq:z_hat}, and grouping together terms of like unit vectors, one can find that:
\begin{equation}
\begin{cases} 
\x=-0.8505 \xp+0.4895 \yp -0.1924 \zp \\
\y=-0.5260 \xp-0.7916 \yp +0.3111 \zp \\
\z=0.3658 \yp + 0.9307 \zp.
\end{cases} 
\end{equation}
This set of equations may be written in matrix form so that the rotation of a vector in the primed frame to the Galactocentric frame is given by the matrix operation
\begin{equation}
\br=\mathscr R^{\prime} \brp,
\end{equation}
where the rotation matrix, $\mathscr R^\prime$, is
\begin{equation}
\mathscr R^\prime = \left[ \begin{matrix} -0.8505 & 0.4895 & -0.1924 \\ -0.5260 & -0.7916 & 0.3111 \\ 0 & 0.3658 & 0.9307\end{matrix} \right].
\end{equation}
The rotation of a position vector in the Galactocentric frame to the primed frame is given as
\begin{equation}
\label{eq:rot_equation}
\brp = \mathscr R \br,
\end{equation}
where, in this case, the rotation matrix $\mathscr R$  is the inverse of $\mathscr R^\prime$ (which, for a rotation matrix, is simply its transpose):
\begin{equation}
\mathscr R = \left[ \begin{matrix} \mathscr R_{11} & \mathscr R_{12} & \mathscr R_{13} \\ \mathscr R_{21} & \mathscr R_{22} & \mathscr R_{23} \\ \mathscr R_{31} & \mathscr R_{32} & \mathscr R_{33} \end{matrix} \right] = \left[ \begin{matrix} -0.8505 & -0.5260 & 0 \\ 0.4895 & -0.7916 & 0.3658 \\ -0.1924 & 0.3111 & 0.9307\end{matrix} \right].
\end{equation}

\section{Converting heliocentric position to center of mass posistion}
\label{sec:convert}

To calculate the kSZ temperature shift due to the diffuse LG medium we choose a line of sight and integrate over $r$.  Since the free electron density, however, is a function of the center of mass distance $R$, we must find a geometric relation between $R$ and a heliocentric position ($r$, $\theta$, $\phi$).  From Fig.~\ref{fig:geometry}, it can be seen that to convert a position vector in the heliocentric frame to the center of mass frame we must (1) translate the vector a distance $d_{\odot}$ to the Galactic center, (2) rotate this vector into the primed coordinate frame, and (3) translate this vector a distance $d_{com}$ to the center of mass origin.

Starting with the most general expression for a position vector expressed in the heliocentric frame, $\bold{r} = (x, y, z)$, the above sequence of transformations can be written mathematically as:
\begin{align}
\label{eq:eq1}
\br & = (\x, \y, \z) \nonumber \\
& = (x-d_{\odot}, y, z),\\
\label{eq:eq2}
\brp & = (\xp, \yp, \zp) \nonumber \\
&= \mathscr R \br \nonumber \\
& = (\x \mathscr R_{11}+\y \mathscr R_{12}, \x \mathscr R_{21}+\y \mathscr R_{22}+\z \mathscr R_{23}, \x \mathscr R_{31}+\y \mathscr R_{32}+\z \mathscr R_{33}), \\
\mathrm{and} \nonumber \\
\bold{R} & = (X, Y, Z)\nonumber \\
& = (\xp, \yp+d_{com}, \zp).
\end{align}
From the previous equation the magnitude of $\bold R$ is
\begin{equation}
R= \sqrt{\mathcalligra{r}^{\: \prime \, 2} +d_{com}^2 +2 d_{com}\yp},
\end{equation}
where $\mathcalligra{r}^{\: \prime \, 2}$ may be re-written in heliocentric variables by noting that $\rp = \rr$ (since they share the same origin):
\begin{equation}
\mathcalligra{r}^{\: \prime \, 2} = \mathcalligra{r}^{\, 2}=(x-d_{\odot})^2+y^2+z^2=r^2+d_\odot^2-2d_\odot x.
\end{equation}
Plugging this into the previous equation and replacing $x$, $y$, $z$ with $r \sin \theta \cos \phi$, $r \sin \theta \sin \phi$ and $r \cos \theta$ respectively, we find an expression for $R$ as a function of $r$, $\theta$ and $\phi$,
\begin{equation}
\label{eq:big_r}
R= \sqrt{r^2+d_\odot^2-2 d_\odot r \sin \theta \cos \phi +d_{com}^2 +2d_{com} \yp (r, \theta, \phi)}.
\end{equation}
The function $\yp (r, \theta, \phi)$ is found from Eqns.~\ref{eq:eq1} and \ref{eq:eq2} and is given as
\begin{align}
\label{eq:yp}
\yp & = (x - d_\odot) \mathscr R_{21}+y \mathscr R_{22}+z \mathscr R_{23} \nonumber \\
& = (r \sin{\theta }\cos{\phi }-d_\odot)\mathscr R_{21}+ r \sin \theta \sin \phi \mathscr R_{22}+ r\cos \theta \mathscr R_{23}.
\end{align}
Eqn.~\ref{eq:big_r} can be inverted to solve for $r$ in terms of $R$ for a certain line of sight, yielding a quadratic, which when solved, results in,
\begin{equation}
\label{eq:r_r}
r(\theta, \phi, R) = -B(\theta, \phi)+\sqrt{B^2(\theta,\phi)+C(R)},
\end{equation}
where the following definitions have been employed:
\begin{equation}
\label{eq:b}
B(\theta, \phi) \equiv d_{com} \mathscr R_{23} \cos \theta+\sin \theta [\cos \phi(d_{com} \mathscr R_{21}  - d_\odot)  +  d_{com} \mathscr R_{22} \sin \phi]
\end{equation}
and
\begin{equation}
\label{eq:c}
C(R) \equiv R^2 - d_{com}^2 - d_\odot^2+ 2 d_{com} d_\odot  \mathscr R_{21}.
\end{equation}

\end{appendices}

\section*{Acknowledgments}
We thank Douglas Finkbeiner and Maxim Lavrentovich for helpful conversations.  This research was supported in part by NSF grant AST-1312034.

\end{document}